\documentclass[runningheads]{llncs}
\usepackage{graphicx}
\usepackage{float}
\usepackage{xcolor}

\usepackage{svg}
\usepackage{subfig}
\usepackage{graphicx}

\usepackage{graphicx}
\usepackage{rotating}
\usepackage{amsmath}
\usepackage{xcolor}
\usepackage{float}

\begin{document}
%


\title{Comparative analysis of the government plans of the Peruvian presidential candidates, SDO(UN) and State Policies of the National Agreement based on NLP}

\titlerunning{Analysis government plans Peruvian presidential candidates}

\author{Honorio Apaza Alanoca\inst{1}, Josimar Chire\inst{2}  \and Jimy Oblitas\inst{3}}

\institute{Data Science Research Group\\ , 
National University of Moquegua, Ilo, Moquegua, Peru \\ 
\and
Institute of Mathematics and Computer Science (ICMC),\\  University of São Paulo (USP), São Carlos, SP, Brazil \\
\and
Facultad de Ingeniería,\\  Universidad Privada del Norte, Cajamarca, Perú \\
\email{hapazaa@unam.edu.pe, jecs89@usp.br, jimy.oblitas@upn.edu.pe} }

\maketitle              
\begin{abstract}
The analysis of government proposal during elections from political parties is vital to choose the next authorities in any city or country. In this paper, we use a text mining approach to analyze the documents and provide an easy visualization to support an easy analysis. Besides, a comparison with a national plan based on sustainable development objectives of UN(United Nations) from 2030 Agenda is perfomed using Natural Language techniques. 

\end{abstract}

\section{Introduction}

Election of authorities is an important event, because citizens will choose the people who will represent them and purpose projects to improve the national, regional context. Traditionally, political parties promote their candidates through mass media, i.e. radio, television, social networks and more. Candidates travel to visit cities and gain more electors.

In Peru, to participate in president elections is a requirement to send a government proposal or plan to Jurado Nacional de Elecciones (National Elections Jury). This document summarizes the proposal of the candidates, considering the most important problems for the party and solutions that they purpose. Usually, these documents have dozens of pages and these are not read for citizens to choose the next authority. Besides, United Nations (UN) purposed an 2030 Agenda to summarize the most important issues which need special attention for governments related to poverty, communication, discrimination and more. 

In 2015, the United Nations (UN) adopted a new international development agenda: the 2030 Agenda that includes the 17 Sustainable Development Goals and 169 targets. This agenda specifies the need for actions to strengthen sustainable economic growth, decent employment and industrialization in all countries\cite{caribbean_2030_2017}.

The 2030 Agenda considers a complex combination of fairly detailed thematic targets, through a comprehensive approach that requires addressing sustainable development as a necessary integration of the social, economic and environmental axes \cite{nieto_crecimiento_2017}. Although it is recognized that each country has its own priorities, this agenda is a reference for government plans seeking an adequate sustainable development of Peru. Therefore, measuring the alignment or possible evolution of government plans of presidential candidates is a necessary task.

In this context, the use of software tools, such as text mining, emerges as a quick and interesting proposal to measure trends. In addition to the fact that, in the Peruvian context, such tools are not used yet, this contrasts with global trends in the use of software tools that are already established, as in the campaigns of Trump and Bolsonaro, in the United States (USA) and Brazil, which illustrate policy facts that have been favored by ICTs \cite{garcia-nunes_computational_2020}.

Natural language processing has shown potential as a promising tool to exploit urban data sources. Authors, such as \cite{cai_natural_2021}, suggest that the use of urban big data sources is still starting and the most studied areas are: urban governance and management, public health, land use and functional zones, mobility and urban design, having been very useful in expanding study scales and reducing research costs.

Text Mining area uses a well-know Data Mining approach, from Data Collection, Exploration, Analysis to Visualization. Text Mining focuses in Text Analysis, uses Natural Language Techniques (NLP). Many studies were performed to analyze different problems from different areas, i.e. epidemiology \cite{chire_saire_study_2020}, politics \cite{sharma_intelligent_2020}, marketing, etc. 

Applications of Text Mining in Politics and Elections, i.e. Anticipating Political Behaviour \cite{sangar2013}, Study Voting Patterns \cite{bagui2007}, Fraud Identification\cite{Poloni2015}, Sentimental Analysis of citizens \cite{SHARMA2020}, Election Result Prediction \cite{Ramteke2016} and more.

The objective of this paper is analyze the government proposal of Peruvian candidates to president elections using a Text Mining Approach to support an easy understanding of the documents. Besides, perform a matching process with national plan adapted from 2030 Agenda, to check how important are these objective for political parties.

Section I includes the review of the bibliography, Section II develops the work proposal, Section III discloses the results of the research and in Section IV gives conclusions, last section presents future work.
\section{Proposal}

Natural language processing is a process transformation the text information in numeric data \cite{DiGiuda2020}. This work is based on the following research process:

\begin{figure}[H]
\centerline{\includegraphics [width=0.9\textwidth]{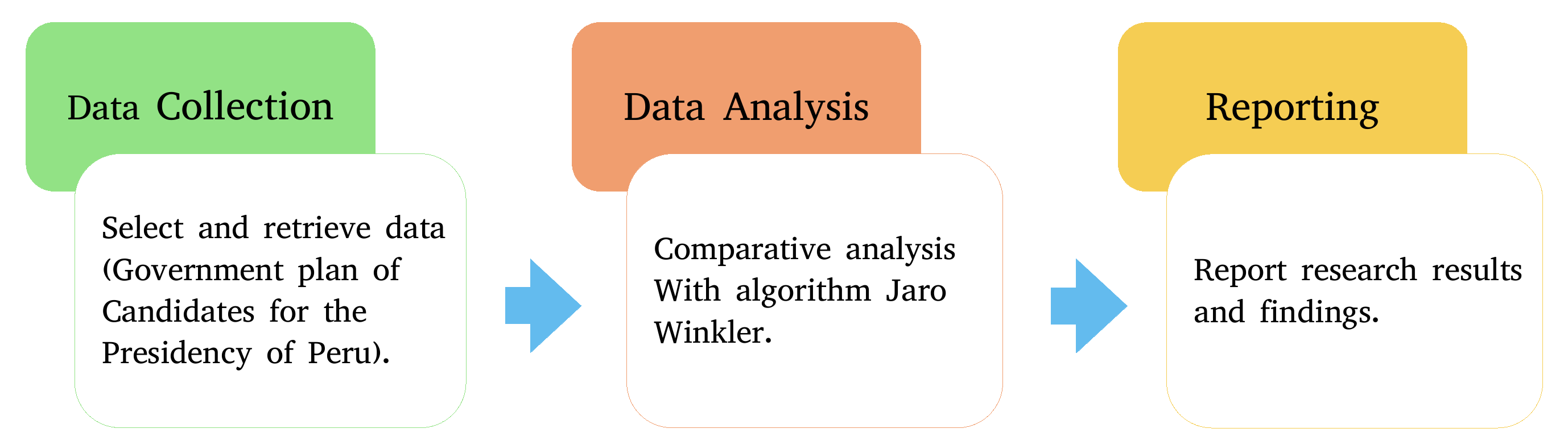}}
\caption{Research process, this process is planed and used for \cite{KIM2017362}}
\label{fig:two}
\end{figure}
\subsection{Data collection}

For the present work, 18 government plans of the candidates for the presidency of the Republic of Peru have been collected. Also the sustainable development goals and policies of the state of the national agreement, the sustainable development goals (SDGs) promoted by the United Nations, whose predecessor are the Millennium Development Goals, constitute an inclusive global agenda with goals for 2030\cite{acuerdonacional}.
\subsection{Data analysis}
Jaro Winkler is the main algorithm to perform comparative text analysis of documents (government plans of the candidates) with the Sustainable Development Goals (SDGs) promoted by the United Nations.
\begin{equation}
Sim_j(s_1, s_2)=\begin{cases} 0 & \text{if } m=0\\\frac{1}{3}(\frac{m}{s_1}+\frac{m}{s_2}+\frac{m-t}{m}) \end{cases}
\end{equation}
The objective is to calculate the distance of the strings of texts that are written in the plans of the government of the candidates and the objectives and policies of sustainable development of the state of the national agreement. In this first preliminary test of the research we are interested in knowing what results are obtained with Jaro Winkler.
\subsection{Reporting}
Finally, the last stage of the research is to make a report on the results obtained, in this case the results are the Jaro Winkler distance between the plans of the candidates' government and the objectives and sustainable development policies of the state of the national agreement.
\section{Results}
This section shows the result frequency of terms in a word cloud, it can be seen that each candidate highlights a particular topic, such as: System, Health, Program, etc. This result is due to the fact that currently the nation and the world are suffering from a global pandemic, therefore, the plans of the candidates' government propose proposals to solve problems related to health. This also shows that other important issues such as education, economics, etc. have been neglected. Especially issues related to sustainable development goals (SDG) promoted by the United Nations.

\begin{figure}[hbpt]
\centerline{\includegraphics [width=0.8\textwidth]{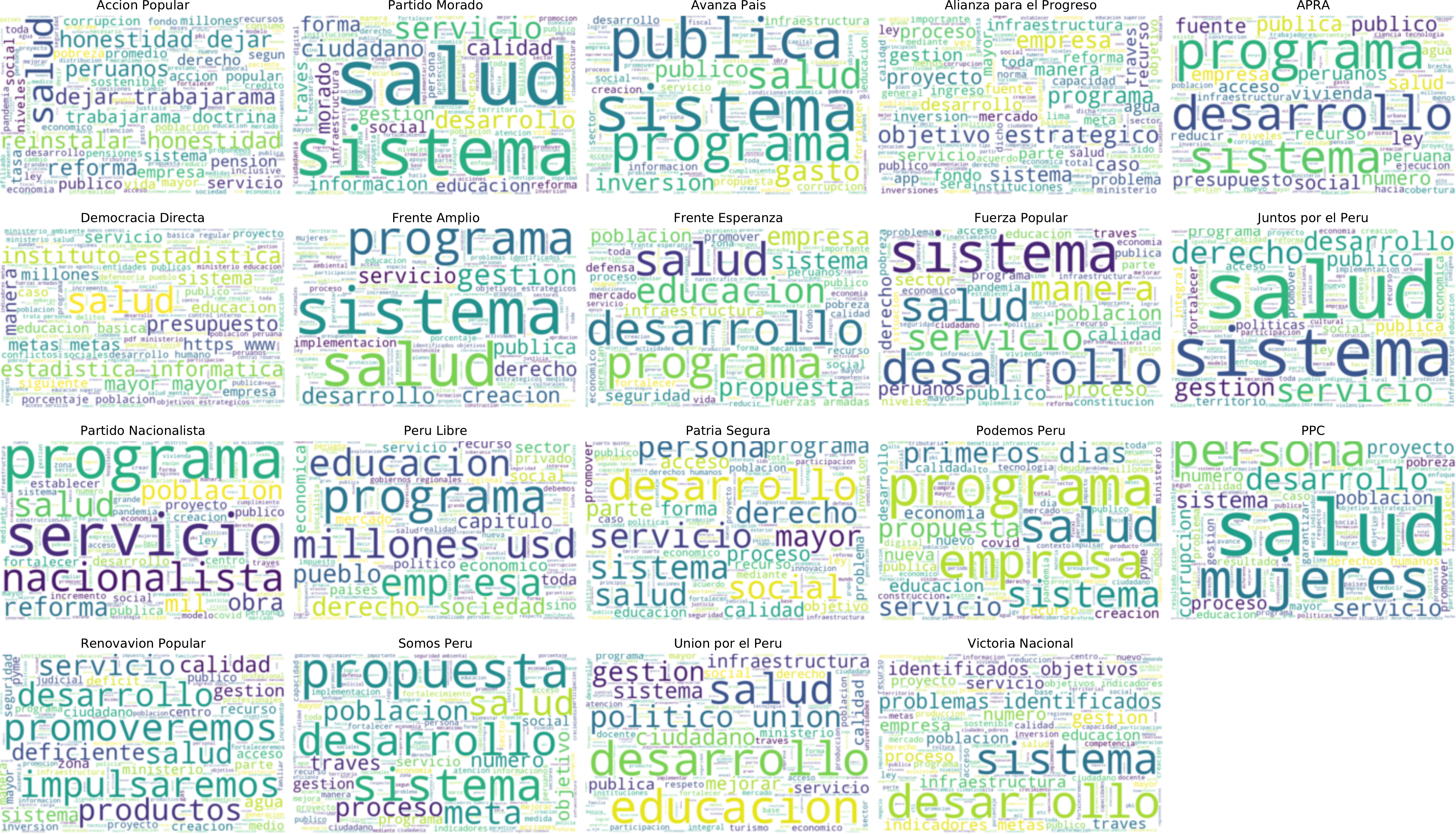}}
\caption{Cloud of words of plans of the government of the candidates}
\label{fig:CW}
\end{figure}
\begin{figure}[hbpt]
\centerline{\includegraphics [width=0.8\textwidth]{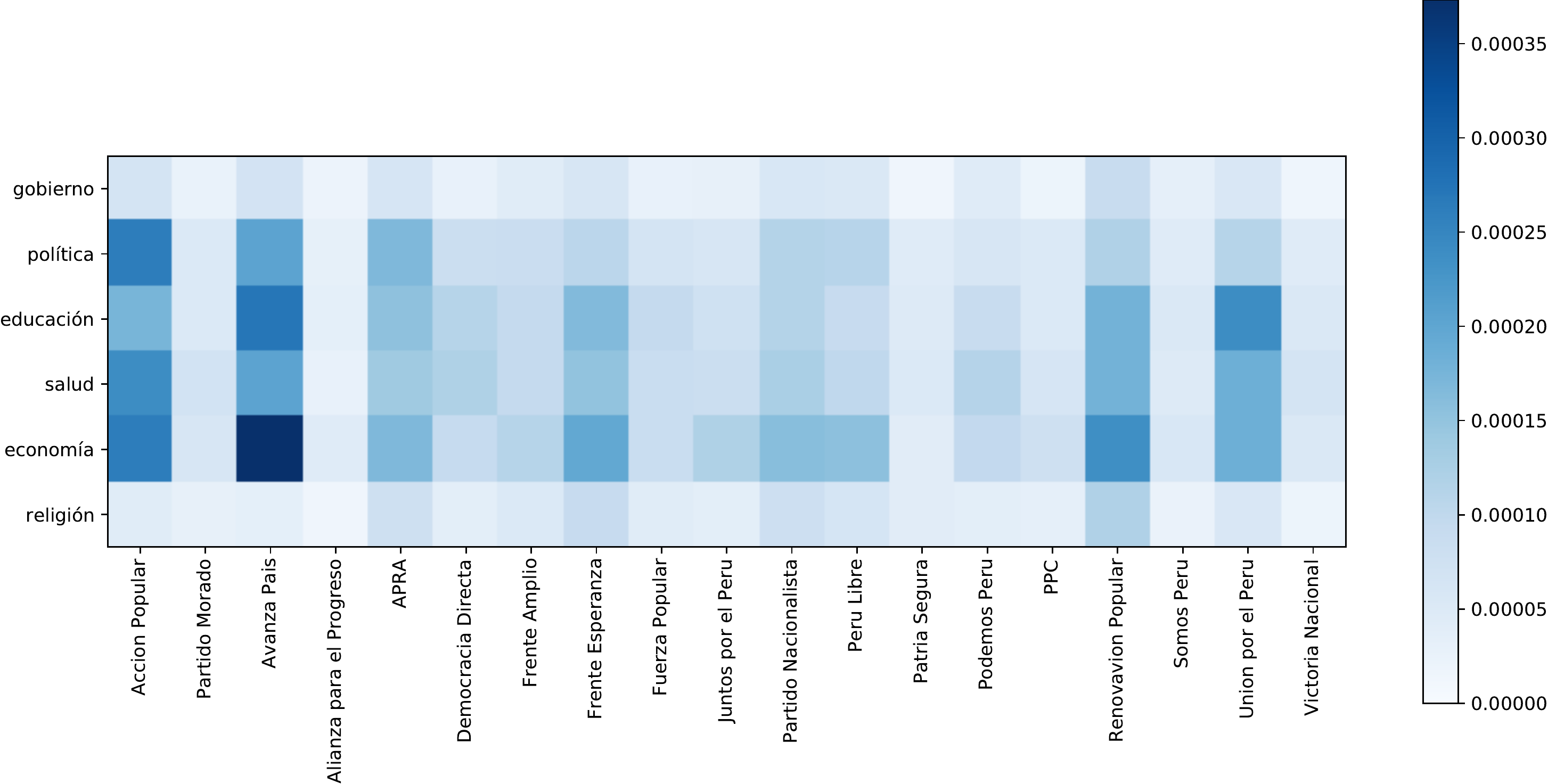}}
\caption{Important areas in the documents}
\label{fig:AR}
\end{figure}

Among the candidates' plans, the one that stands out the most is the government plan of the political party Avanza Pais on the economic issue, It can also be seen that the Accion Popular political party has a uniform distribution in its government plan on the issues of economy, health, education and politics. Can be seen in Figure \ref{fig:AR}.

\begin{figure}[H]
\centerline{\includegraphics [width=0.9\textwidth]{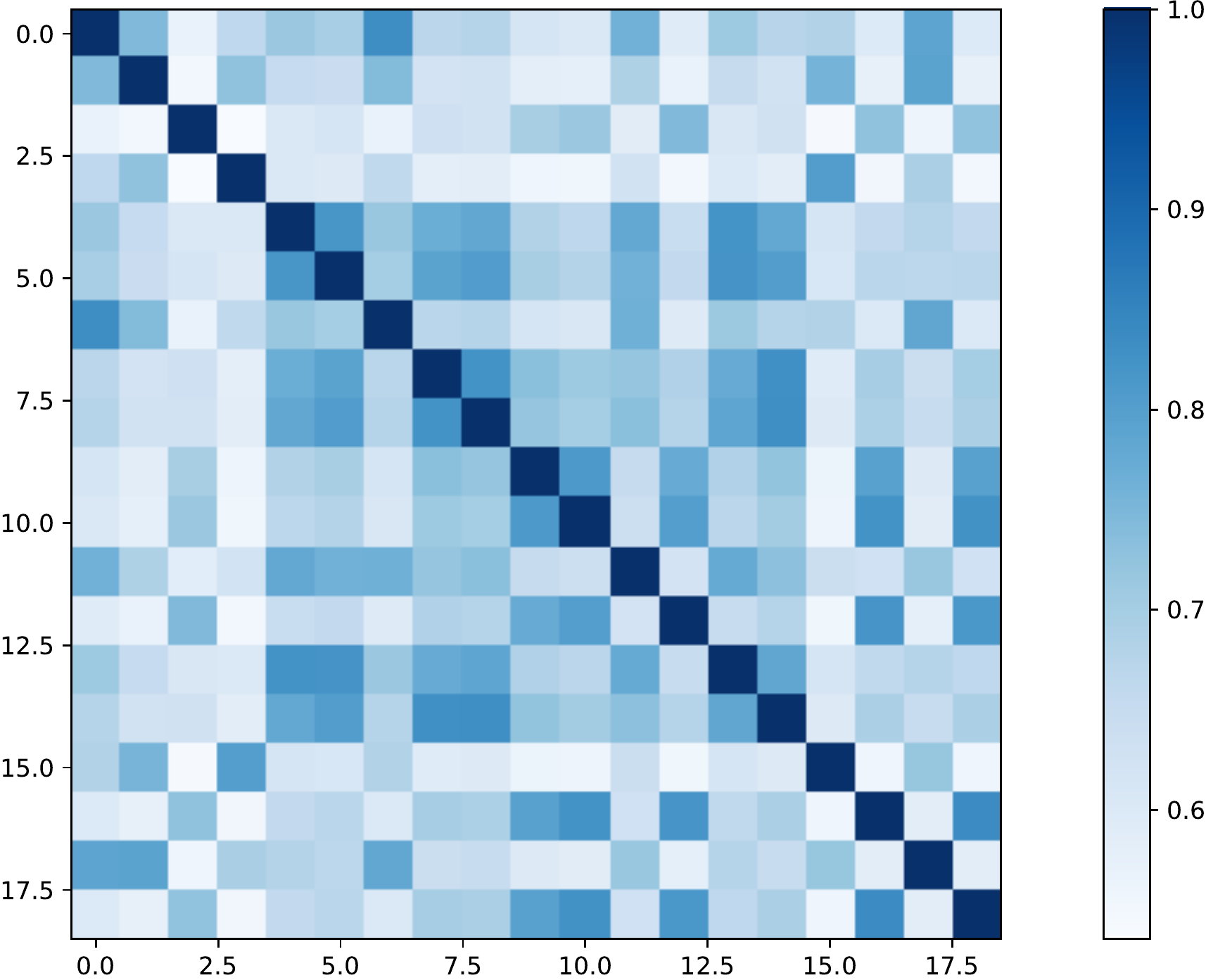}}
\caption{Documents similarity}
\label{fig:COM}
\end{figure}

In this case we can vary the issues we want to measure, this can be according to the context of the moment and different sectors of society, they have different problems and needs, so it is important to analyze from other points of view, social classes and thoughts.\\
Below we present a graphical (Figure \ref{fig:COM}) representation of how similar are the government plans of the candidates for the presidency of Peru, in Figure \ref{fig:COM} it can be seen that they are not so identical, but if you can see the degree of similarity they have, but This is due to the fact that government plans clearly address very similar issues that translate into social problems (health, economy, programs, etc.) and government (judiciary, corruption, congress, etc.).\\
In the experiment, the differences by prolific class were also denoted, in some cases the distance is very noticeable between the political parties considered to be on the left with those on the right. Which can be similar in the daily exercise, which obviously have very different thoughts, therefore very different proposals between these two sides of Peruvian politics.
In this section we are going to analyze the distance between chains of texts written in the government plans of the candidates and the objectives and policies of sustainable development of the state of the national agreement, we try to differentiate the similarities between these two documents, when a chain of texts is similar to another means that the document contains texts similar to the other. So, we could say that a government plan addresses one or many sustainable development goals and policies of the state of the national agreement.

\begin{figure}[H]
\centerline{\includegraphics [width=0.9\textwidth]{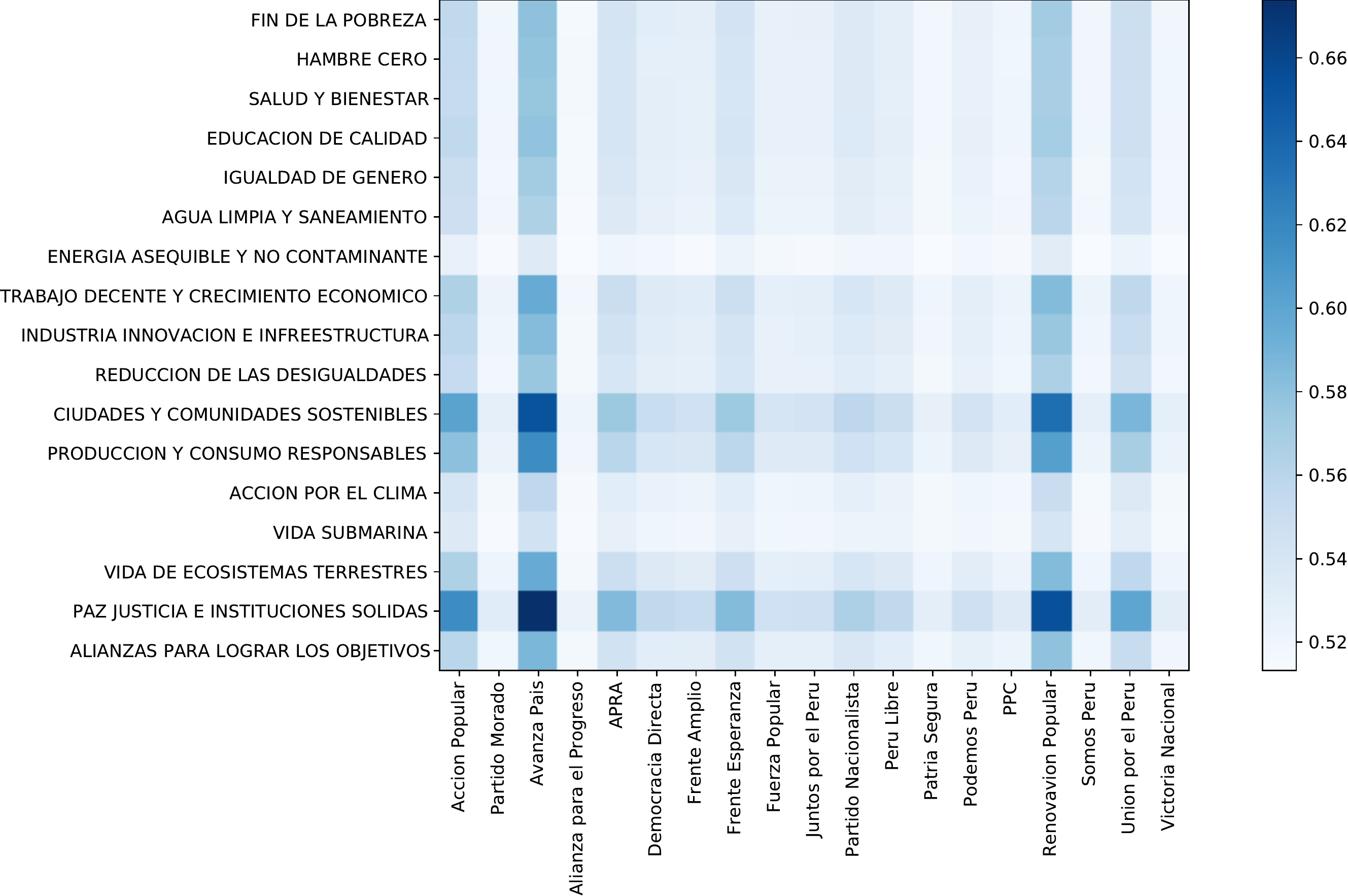}}
\caption{Documents similarity Plan}
\label{fig:two}
\end{figure}

In the graph above, it can be seen that the government plan of the political party Avanza Pais addresses much more than others the goal of peace, justice and solid institutions (paz justicia e instituciones solidas), followed by the political party Renovacion Ppular. However, little is addressed the objectives such as: underwater life(Vida submarina), health and well-being(salud y bienestar), end of poverty(fin de la probreza), etc.




\section{Conclusions}

The algorithm Jaro Winkler based on measuring the distance of text chains shows us that we are very interesting preliminary results, it shows us some differences between the government plans of the candidates for the presidency of Peru, as well as the objectives of the Sustainable Development Goals and the State Policies of the National Agreement. However, these results can be further refined with the most advanced artificial intelligence methods or algorithms.\\
In the present we want to highlight the way in which the differences between the government plan documents can be graphically demonstrated, this way of showing the document differences is very important for the electorate, because without having to read all the government plans, they can obtain a more general vision graphically.
\section{Future work}
One of the future jobs is to experiment with highly advanced artificial intelligence techniques in the discipline of natural language processing and text mining.\\
It would be very interesting to study and experience how coherent the arguments of the candidates are in the debate with their government plan. Because there must be coherence of ideas between the proposals that are written in the government plan with what the candidate expresses in the debate, interviews in the press, etc.



\bibliographystyle{apalike}
\bibliography{biblio.bib}

\end{document}